\documentclass[preprint2]{aastex61}
\usepackage{amsmath}
\usepackage{xcolor}

\def\HII{{\ion{H}{2}}}
\def\OII{[{\ion{O}{2}}]}
\def\4959_5007{[\ion{O}{3}]~$\lambda \lambda$4959,5007}
\def\OIII49595007{[\ion{O}{3}]~$\lambda \lambda 4959,5007$}
\def\ratioR23{([\ion{O}{2}]~$\lambda \lambda$3727,9 + [\ion{O}{3}]~$\lambda\lambda$4959,5007)/H$\beta$}
\def\R23{${\rm R}_{23}$}
\def\dS23{${\rm S}_{23}$}

\def\NII{[{\ion{N}{2}}]}

\def\NIIOII{[\ion{N}{2}]/[\ion{O}{2}]}

\def\ratioS23{([\ion{S}{2}]~$\lambda \lambda$6717,31 +[\ion{S}{3}]~$\lambda\lambda$9069,9532)/H$\beta$}
\def\NIIHa{[\ion{N}{2}]/H$\alpha$}

\def\SII{[{\ion{S}{2}}]}

\def\Hb{{H$\beta$}}
\def\O4363{[{\ion{O}{3}}]~$\lambda$4363}
\def\OIII{[{\ion{O}{3}}]}
\def\OIIIHb{[\ion{O}{3}]/H$\beta$}
\def\Ha{{H$\alpha$}}
\def\Ep{$E_{\rm peak}$}
\def\fN{$f_{\rm NLR}$}

\shorttitle{AGN-HII mixing in SDSS}
\shortauthors{A. D. Thomas et al.}

\begin{document}

\title{Mixing between Seyfert and \HII-region excitation in local active galaxies}

\correspondingauthor{Adam D. Thomas}
\email{adam.thomas@anu.edu.au}

\author[0000-0003-1761-6533]{Adam D. Thomas}
\author{Lisa J. Kewley}
\author{Michael A. Dopita}
\author{Brent A. Groves}
\affiliation{RSAA, Australian National University, Cotter Road, Weston Creek, ACT 2611, Australia}
\affiliation{ARC Centre of Excellence for All Sky Astrophysics in 3 Dimensions (ASTRO 3D), Cotter Road, Weston Creek, ACT 2611, Australia}

\author{Andrew M. Hopkins}
\affiliation{Australian Astronomical Observatory, 105 Delhi Rd, North Ryde, NSW 2113, Australia}
\affiliation{ARC Centre of Excellence for All Sky Astrophysics in 3 Dimensions (ASTRO 3D), 105 Delhi Rd, North Ryde, NSW 2113, Australia}

\author{Ralph S. Sutherland}
\affiliation{RSAA, Australian National University, Cotter Road, Weston Creek, ACT 2611, Australia}
\affiliation{ARC Centre of Excellence for All Sky Astrophysics in 3 Dimensions (ASTRO 3D), Cotter Road, Weston Creek, ACT 2611, Australia}

\begin{abstract}
We fit theoretical models to the emission-line spectra of 2766 Seyfert galaxies from the Sloan Digital Sky Survey (SDSS).  The model line fluxes are derived by `mixing' photoionization model predictions of active galactic nucleus (AGN) narrow line region (NLR) emission and \HII\ region emission.  The observed line fluxes in each spectrum were directly compared to the grid of mixed models using the Bayesian parameter estimation code NebulaBayes, thereby measuring the degree of mixing in each spectrum for the first time.  We find that the majority of the Balmer line emission in the majority of Seyfert-classified SDSS spectra arises from contaminating \HII\ regions within the fixed-size aperture.  Even for spectra with $\log$\,\OIIIHb$\;\gtrsim 0.9$, a median of ${\sim}30\%$ of the Balmer flux arises in \HII\ regions.  We show that the results are qualitatively insensitive to the assumed Seyfert ionizing continuum, and that ionizing spectra with a peak energy of \Ep$\,\sim 40$ -- 50~eV produce the most plausible distributions of mixing fractions.  The analysis cleanly quantifies how the starburst -- AGN `mixing fraction' increases on the BPT diagram for SDSS galaxies.  Apart from the mixing fraction, the models also vary with the ionization parameter in the NLR, the gas pressure, and the metallicity.  Measurements for the metallicity in particular will be presented in a companion paper.
\end{abstract}

\keywords{Galaxies: active, Galaxies: emission lines, Galaxies: ISM, Galaxies: Seyfert}

\section{Introduction} \label{sec:Intro}

The Sloan Digital Sky Survey \citep[SDSS;][]{York_2000_SDSS} provides a rich spectroscopic sample of galaxies in the local universe, including tens of thousands of galaxies hosting Active Galactic Nuclei (AGN).  The SDSS AGN population has been extensively studied, providing insights into many aspects of the population: the host galaxy properties \citep[e.g.][]{Kauffmann_2003, Kewley2006_AGN_hosts, 2011_Koss_BAT_SDSS_AGN, 2015_Reines_BH_mass}, nuclear activity as a function of galaxy environment \citep[e.g.][]{Miller_2003_SDSS_AGN, Constantin_2008_AGN_in_voids, von_der_Linden_2010_SDSS_clusters, Khabiboulline_2014_SDSS_AGN} classification of excitation mechanisms \citep[e.g.][]{Kewley2006_AGN_hosts, Shirazi_Brinchmann_2012_HeII, Bar_2017_AGN_diagnostic}, gas kinematics \citep[e.g.][]{Greene_2005_SDSS_AGN_kinematics}, comparisons to AGN emission at other wavelengths \citep[e.g.][]{Kauffmann_2008_SDSS_radio_AGN, 2011_Koss_BAT_SDSS_AGN, 2013_Mateos_AGN, Paliya_2018_SDSS_AGN}, relative alignment of the host galaxy and active nucleus \citep{Lagos_2011_SDSS_AGN_alignment}, and the distribution of black hole masses and Eddington ratios \citep[e.g][]{Greene_2007_SMBH_mass_function, Kauffmann_2009_Edd_ratio, Jones_2016_SDSS_LEdd} amongst other topics.

A large proportion of SDSS AGN spectra show clear mixing between \HII\ region emission and AGN NLR emission on optical diagnostic diagrams.  This mixing is broadly attributable to the summing of emission in the SDSS fiber aperture, but more subtly is associated with the correlation between star formation rate and population-averaged supermassive black hole accretion rate \citep[e.g.][]{Chen_2013_SFR_vs_BHAR} and with the close timing of starbursts and black hole accretion \citep[e.g.][]{Wild_2010_SB_AGN_connection}.  Although integral field data now allows detailed spatial analysis of the mixing of \HII\ and NLR emission \citep[e.g.][]{Davies_2014_Mixing_2, Davies_2014_Mixing_1, Dopita_2014_probing_ENLR_I_S7}, the effect of \HII\ region contamination on single-fibre SDSS NLR spectra has not been robustly quantified.  The \HII\ -- NLR mixing significantly complicates detailed photoionization modelling of the SDSS AGN.

In this Letter we tackle the \HII\ -- NLR mixing using theoretical models and a general Bayesian method, determining the proportion of the emission arising from each of the two mechanisms for each SDSS AGN spectrum.  We demonstrate that the mixing is an important consideration even for the highest-excitation AGN spectra in the SDSS sample.

\section{Observational data} \label{sec:Obs_data}

Our sample is taken from SDSS DR7 \citep{Abazajian_2009_SDSS7}, and we use the emission line fluxes provided by the MPA/JHU catalogue\footnote{\url{www.mpa-garching.mpg.de/SDSS/DR7/}\\ \url{http://home.strw.leidenuniv.nl/~jarle/SDSS/}} \citep{Tremonti_2004_MZ}.  We selected galaxies with redshifts in the range $0.02 < z < 0.37$, which resulted in the g-band covering fraction (proportion of the galaxy light within the fiber aperture) having a 25th percentile, median, and 75th percentile of 0.15, 0.22 and 0.30, respectively.  The lower redshift cut involves a trade-off between sample size and potential systematic aperture effects due to low covering fractions \citep{Kewley_2005_Aperture_effects}, and is also determined by the need for the \OII\ doublet to be redshifted into the observed wavelength range.

We applied the following further cuts:
\begin{enumerate}
    \item All fluxes and errors of relevant emission lines were required to be finite and positive
    \item A signal to noise (S/N) cut was applied based on the \Hb\ flux.  We required \Hb\ ${\rm S/N} > 5$ and did not use S/N cuts on other lines to avoid biasing the sample (in particular, biasing the metallicities).
    \item We required spectra to have a Balmer decrement \Ha\ / \Hb\ greater than 2.7, because lower values are unphysical, and we require reliable flux calibration and reddening corrections.
\end{enumerate}

After these cuts there remained 2766 Seyfert-classified objects and 51806 \HII-classified objects.  The spectra were classified using the \NII\ and \SII\ optical diagnostic diagrams \citep{BPT1981, Veilleux_1987_LineRatios}, following the method of \citet{Kewley2006_AGN_hosts}.  Only the Seyfert- and `composite'-classified galaxies were included in the analysis, with LINER, star-forming, and ambiguous galaxies being excluded.

\section{Method} \label{sec:Method}

The primary tool we use in our analysis is NebulaBayes\footnote{\url{https://github.com/ADThomas-astro/NebulaBayes}}, a package that performs Bayesian parameter estimation by comparing observed emission line fluxes and errors with an arbitrary grid of model fluxes.  NebulaBayes is described in detail by \citet{Thomas_2018_NebulaBayes}.

The MAPPINGS~V \citep{Sutherland_Dopita_2017_shocks_MAPPINGSV} photoionization model grids we feed into NebulaBayes are `mixing' grids, derived by combining \HII\ region and NLR model grids.  Both grids have dimensions in the parameters of oxygen abundance $12 + \log {\rm O/H}$, ionization parameter at the inner edge of the photoionization model $\log U$, and gas pressure $\log P/k$.  In addition, the NLR grid has a fourth parameter \Ep, which is the peak (on a log-log plot of energy flux versus energy) of the `big blue bump' ionizing accretion disk emission in the model of the ionizing spectrum \citep{Thomas_2016_oxaf}.  The \HII\ and NLR grids used here are the same as those distributed with the NebulaBayes software \citep{Thomas_2018_NebulaBayes}, with the exception of four intermediate \Ep\ values that have been added to the NLR grid.

The \HII\ and NLR grids were combined into a `mixing' grid using the following assumptions:
\begin{enumerate}
	\item The \HII\ regions and NLR clouds within the same aperture have the same metallicity
	\item The \HII\ regions and NLR clouds within the same aperture have the same pressure, $\log P/k$
	\item There is little sensitivity to the \HII\ region ionization parameter.  This parameter was fixed at $\log U_{\rm HII} = -3.25$, a representative value for high metallicity galaxies that was found by analysing the SDSS \HII\ spectra using NebulaBayes with the \HII\ region grid alone.
	\item \Ep\ can be fixed to a single representative value without biasing the results.
\end{enumerate}

The number of parameters is limited by computational practicalities and by the number of independent line fluxes in each spectrum.  A more complex model is unlikely to be justified by statistically significant improvements to model fits.

The `mixing' is parametrized by \fN, the proportion of the \Hb\ flux in a `mixed' spectrum that arises from NLR as opposed to \HII\ emission ($f_{\rm HII} = 1 - f_{\rm NLR}$).  This parametrization, first used by \citet{Kewley_2001_optical_classification_IR_galaxies}, is more sophisticated than approaches that use distances on the optical diagnostic diagrams to parametrize the gas excitation \citep[e.g.][]{Khabiboulline_2014_SDSS_AGN}.  We sampled \fN\ at the six values $f_{\rm NLR} = 0.0$, 0.2, 0.4, 0.6, 0.8 and 1.0.  In total, there were four parameters in each mixing grid (with a different grid for each considered \Ep\ value): $12 + \log {\rm O/H}$, $\log P/k$, $\log U_{\rm NLR}$ and \fN.

The following emission lines were used in the analysis: \OII$\,\lambda\,3726 + 29$, [\ion{Ne}{3}]$\,\lambda\,3869$, \OIII$\,\lambda\,4363$, \OIII$\,\lambda\,5007$, \ion{He}{1}$\,\lambda\,5876$, [\ion{O}{1}]$\,\lambda\,6300$, \NII$\,\lambda\,6583$, and \SII$\,\lambda\,6716 + 31$.  We also included \Ha\ and \Hb, which are essential for both normalization and reddening-correction.  We used the built-in ability of NebulaBayes to deredden spectra to match the predicted Balmer decrement at every point in the model grid.

Two priors were combined, with equal weighting.  These were a prior on the \SII\ doublet flux ratio, to constrain the gas pressure, and a prior on the \NIIOII\ ratio, to help constrain the metallicity.

\section{Results} \label{sec:results}

\subsection{Variation in the mixing fraction} \label{sec:BPTVO_fNLR}

\begin{figure*}[hbtp]
	\centering
	\includegraphics[width=1.0\textwidth]{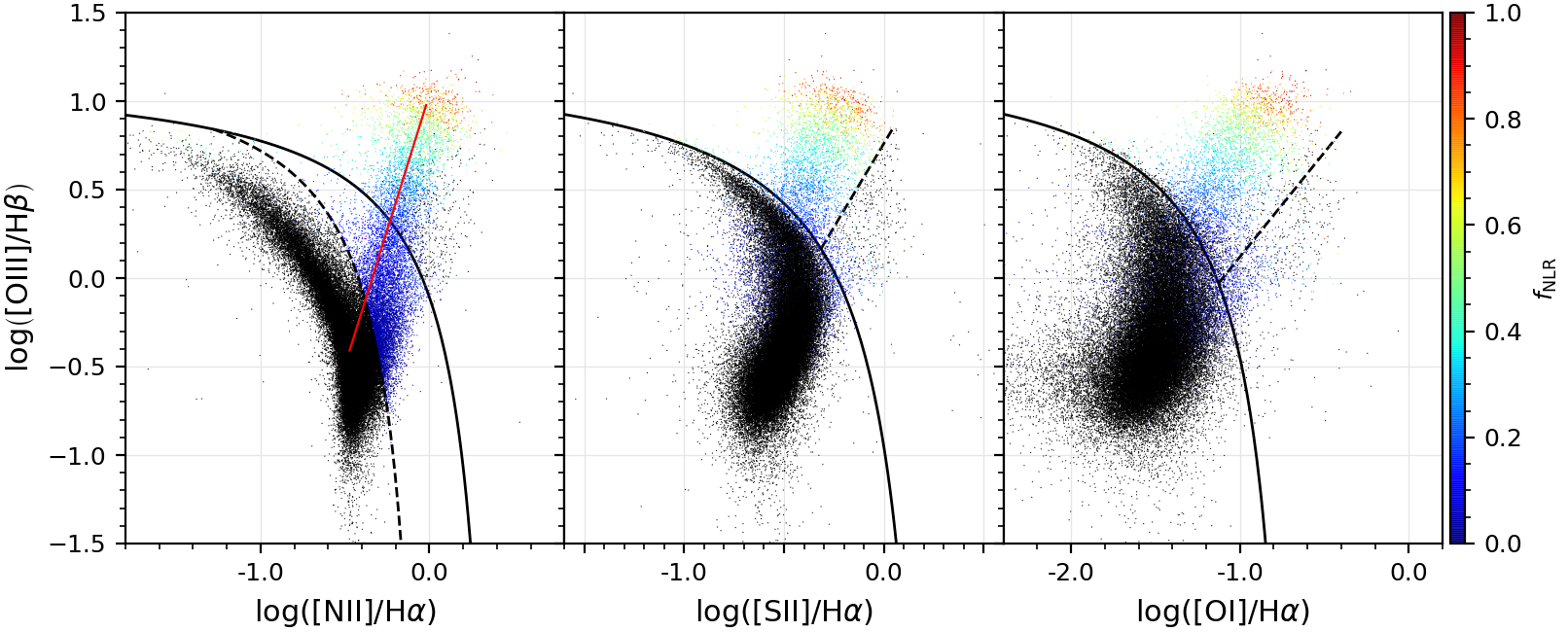}
	\caption{Optical diagnostic diagrams showing SDSS data colored by the fitted values of \fN, the proportion of the \Hb\ flux that arises in NLR as opposed to \HII-region emission.  The colored points are classified as `composite' or Seyfert using the \NII\ and \SII\ diagrams, with all other classifications shown in black.  The dashed line \citep{Kauffmann_2003} in the left-most panel gives an empirical upper-limit on the line ratios of star-forming galaxies, and the solid black lines in all three panels give corresponding theoretical limits \citep{Kewley2001}.  The dashed lines in the rightmost two panels approximately separate Seyferts (above) from LINERs \cite[below;][]{Kewley2006_AGN_hosts}, and the red line in the leftmost panel is chosen to approximately follow the \HII-Seyfert `mixing sequence'.  There is a clear increase in \fN\ up the mixing sequence from star-forming galaxies to Seyfert galaxies.  This analysis used a Seyfert ionizing spectrum with \Ep$ \, = 45$\,eV.  \label{fig:BPTVO_3panel}\\}
\end{figure*}

Our results for the parameter \fN\ are presented in Figure~\ref{fig:BPTVO_3panel} on optical diagnostic diagrams, with the classification lines described by \citet{Kewley2006_AGN_hosts} shown for reference.  The results show that the mixing fraction \fN\ behaves intuitively by increasing upwards along the mixing sequence from the star-forming sequence to the Seyfert region.

The red line in the left panel of Figure~\ref{fig:BPTVO_3panel} is a `mixing line', which is chosen to be linear in log-log space and follow the upper locus of the \HII-AGN mixing sequence.  The line connects the median \HII-classified point (at $(x,y) = (-0.466, -0.408)$) and the median Seyfert-classified point with $y > 0.9$ (at $(x,y) = (0.003, 0.979)$), where $x = \log\;$\NIIHa\ and $y = \log\;$\OIIIHb.  We note that a true `mixing locus' would be curved in the Figure~\ref{fig:BPTVO_3panel} log-log plot.

\subsection{Effect of the ionizing Seyfert spectrum} \label{sec:E_peak}

The mixing fraction \fN\ is degenerate with the parameter corresponding to the hardness of the Seyfert radiation field, \Ep.  Increasing either quantity has the effect of increasing the effective ionization of a `mixed' model spectrum.  Understanding the effect of \Ep\ allows us to quantify typical \fN\ values as a function of the position of SDSS galaxies on the optical diagnostic diagram.

\begin{figure*}[htbp]
	\centering
	\includegraphics[width=0.85\textwidth]{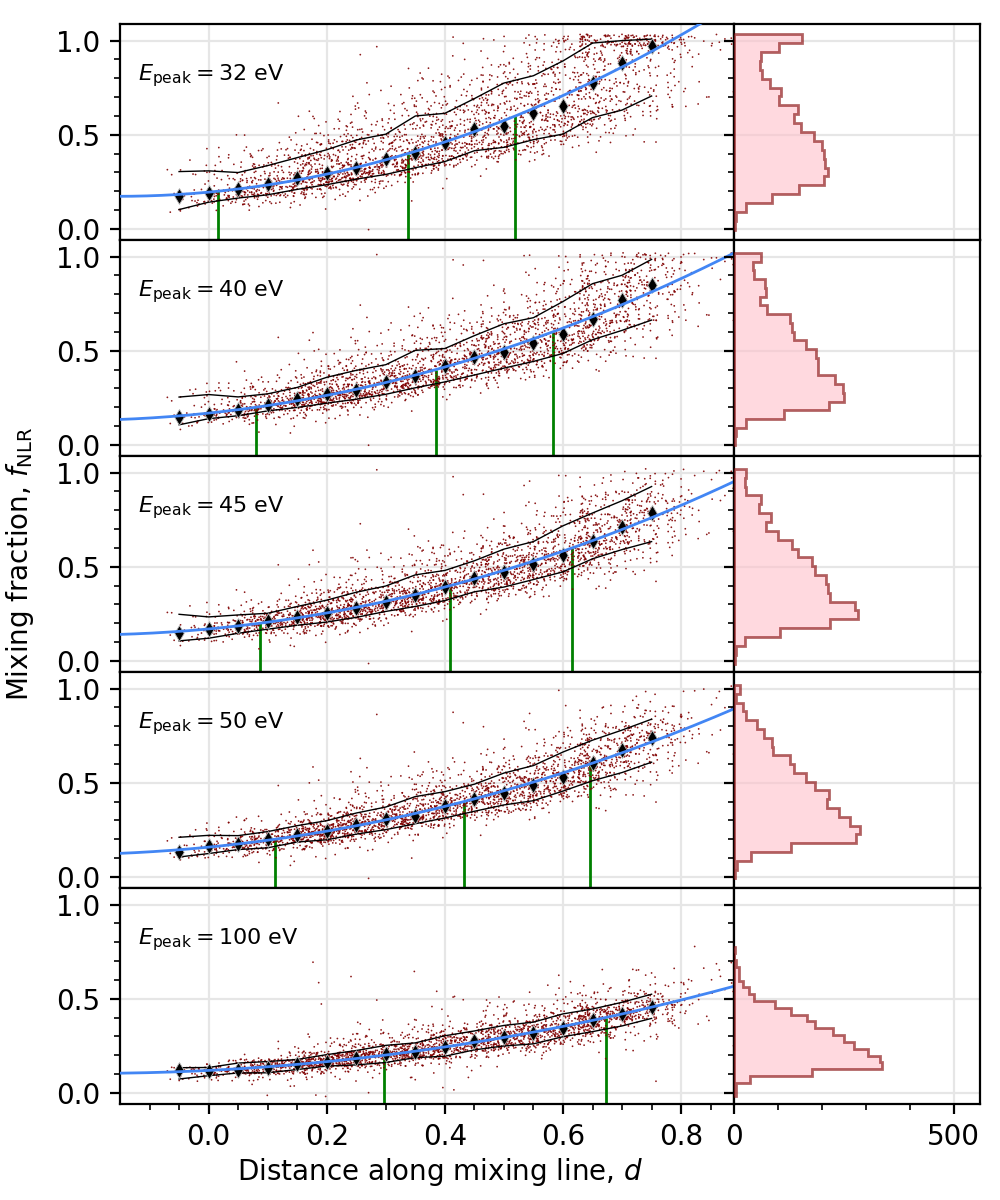}
	\caption{The inferred mixing fraction \fN\ versus $d$, the projected distance along the `mixing line' (the red line in Figure~\ref{fig:BPTVO_3panel}; the intersection with the \citet{Kewley2001} line defines $d=0$). Results are presented for Seyfert-classified galaxies for five values of \Ep, the peak of the ionizing Seyfert continuum spectrum.  Black diamonds show the median in each $d$-bin, and black lines show the 16th and 84th percentiles.  The blue curve is a parabola fitted to the medians and errors.  Green vertical lines indicate where 20, 40 and 60\% of the Balmer flux arises due to Seyfert photoionization as opposed to \HII-region emission, according to the blue fits.  For all plausible values of \Ep, the analysis demonstrates that the majority of the line emission in the majority of Seyfert-classified SDSS galaxies arises in \HII\ regions as opposed to the NLRs.  The parameters \Ep\ and \fN\ both vary the ionization level of the models and therefore are difficult to disentangle, but the figure shows that the inferred \fN\ sequence is qualitatively insensitive to \Ep, with \Ep$\, \sim 40 - 50$\,eV producing a plausible \fN\ distribution.  \label{fig:fNLR_vs_mix_dist}}
\end{figure*}

Figure~\ref{fig:fNLR_vs_mix_dist} shows the inferred mixing fraction \fN\ versus projected distance along the mixing line $d$, for a series of \Ep\ values.  The $d = 0$ point was chosen to coincide with the intersection of the red mixing line and the solid black line \citep[extreme starburst line of][]{Kewley2001} in Figure~\ref{fig:BPTVO_3panel}.  For \Ep\,$ = 32$~eV the high-\fN\ data is clustered at \fN\,$ = 1$, presumably because the observed spectra have a higher ionization than the models in the relevant part of the parameter space, resulting in the inferred \fN\ being artificially elevated.  Conversely, at the other extreme of \Ep\,$ = 100$~eV (improbably high considering the ionization potentials of species observed in typical NLRs), all galaxies have a relatively low inferred \fN\ value.  Results do not change significantly for \Ep\ in the range \Ep$\,\sim 40 - 50$~eV, and we select \Ep\,$ = 45$\,eV to use in quantifying typical mixing fractions on the mixing sequence.

The scatter in \fN\ may be associated with variations in the three other parameters, which would allow a range of \fN\ values to correspond to a single $d$ value.  However, a major contribution to the scatter is likely to be variation in the intrinsic \Ep\ values of the AGN population.  We expect \Ep\ to increase with the Eddington ratio $\lambda_{\rm Edd}$, and the SDSS Seyfert $\lambda_{\rm Edd}$ distribution is known to have a non-trivial width \citep[e.g.][]{Kewley2006_AGN_hosts, Kauffmann_2009_Edd_ratio}.  Variation in $\lambda_{\rm Edd}$ should cause more scatter at high \fN, and indeed the scatter does noticeably increase with \fN\ (and $d$) in all panels of Figure~\ref{fig:fNLR_vs_mix_dist}.

\subsection{Measuring typical mixing fractions} \label{sec:quantifying_fNLR}

\begin{figure}
	\centering
	\includegraphics[width=0.48\textwidth]{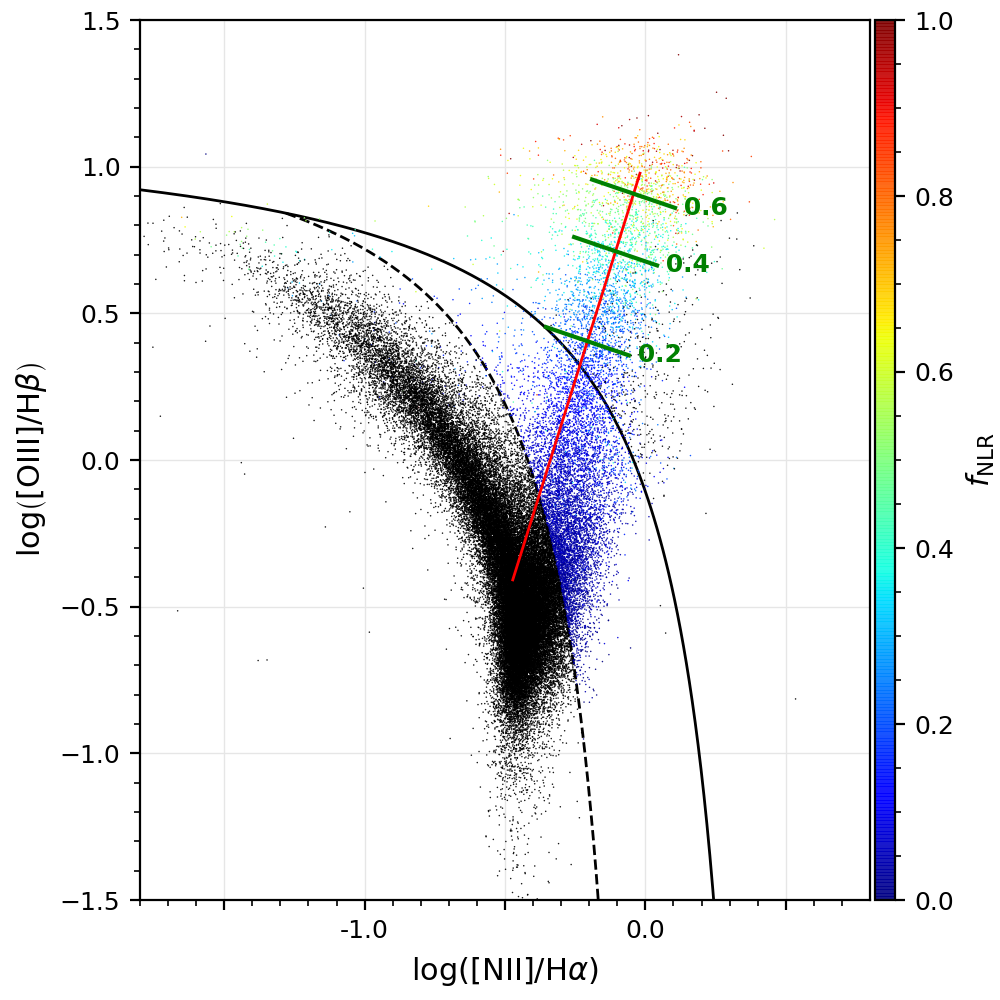}
	\caption{Optical diagnostic diagram as in Figure~\ref{fig:BPTVO_3panel}, but with green lines added to indicate the median mixing fraction \fN\ on the \HII\ to AGN mixing sequence.  The \fN\ estimates were obtained using the `mixing' model grid with \Ep\,$ = 45$~eV.\\  \label{fig:BPT_NII_results}}
\end{figure}

We now consider the typical mixing fractions as a function of location on the BPT diagram.  Figure~\ref{fig:BPT_NII_results} shows an \NII\ optical diagnostic diagram, illustrating the derived lines of `median mixing fraction'.  These green mixing fraction lines correspond to the green vertical lines in the \Ep$ = 45$~eV panel of Figure~\ref{fig:fNLR_vs_mix_dist}.  In Figure~\ref{fig:BPT_NII_results} these lines are drawn perpendicular to the red mixing line, and have a gradient of $m = -0.327$.  The intercept $b$ (where $y = mx + b$) for each line is 0.341, 0.676, and 0.893 for \fN$ = 0.2$, 0.4 and 0.6 respectively.

A key result evident in Figures~\ref{fig:fNLR_vs_mix_dist} and \ref{fig:BPT_NII_results} is that for the majority of Seyfert-classified SDSS spectra, the majority of the Balmer emission arises in \HII\ regions, and not in the NLRs.  Only spectra with $\log$\,\OIIIHb$\, \gtrsim 0.9$ typically have the majority of the Balmer flux arising in the Seyfert NLRs.  These results hold true for all \Ep\ in the range $40 - 50$\,eV.

\section{Discussion} \label{sec:Discussion}

We have clearly demonstrated that modeling of Seyfert emission in the SDSS sample must take \HII-region contamination into account.  Cuts on BPT classifications may remove AGN spectra with a high degree of \HII\ contamination \citep[e.g.][]{Maragkoudakis_2014_BPT_aperture_effects}, but our results show that a large fraction of the remaining spectra are significantly contaminated.  A small proportion of LINER galaxies may remain in our Seyfert sample due to the approximate nature of the Seyfert-LINER separation \citep{Kewley2006_AGN_hosts}.  However, these interloping LINERs are unlikely to affect the results (especially considering that ionization parameter varies in our models).

Table~\ref{table:flux_fractions} shows the relative contributions of NLR emission and \HII\ region emission to line fluxes in mixed model spectra.  The \OIII\ emission is dominated by NLR clouds even at low \fN, but the proportional NLR flux contribution is consistently below \fN\ for \NII\ and \SII.  Evidently \OIII\ is a highly sensitive tracer of AGN contamination in \HII\ spectra.  Conversely, for other lines, the Seyfert-classified spectra may easily suffer significant contamination from \HII-region emission.  The data in Table~\ref{table:flux_fractions} appears to be consistent with Figure~3 of \citet{Kauffmann_2009_Edd_ratio}, who also calculate likely \HII\ region contributions to the \OIII\ flux.

\begin{deluxetable*}{lcccccc}[htbp]
	\centering
	\tabletypesize{\scriptsize}
	\tablewidth{0.8\textwidth}
	\tablecaption{Proportion of the flux in a `mixed' spectrum that is from AGN NLR models as opposed to \HII\ region models, for each strong optical line.  The models have $12 + \log {\rm O/H} = 8.990$, $E_{\rm peak} = 45$~eV, $\log U_{\rm NLR} = -2.60$, $\log U_{\rm HII} = -3.25$, and $\log P/k \, ({\rm cm^{-3}\,K}) = 7.40$.\label{table:flux_fractions}}
	\tablehead{
		$f_\mathrm{NLR}$ & H$\beta$ & [OII]$\lambda\lambda 3726,29$ & [OIII]5007 & [OI]6300 & [NII]6583 & [SII]$\lambda\lambda 6716,31$
	}
	\startdata
	0.0   &    0.0 &       0.000 &     0.000 &   0.000 &    0.000 &       0.000 \\
	0.1   &    0.1 &       0.213 &     0.952 &   0.403 &    0.092 &       0.093 \\
	0.2   &    0.2 &       0.378 &     0.978 &   0.603 &    0.186 &       0.188 \\
	0.4   &    0.4 &       0.618 &     0.992 &   0.802 &    0.379 &       0.381 \\
	0.6   &    0.6 &       0.785 &     0.996 &   0.901 &    0.579 &       0.581 \\
	0.8   &    0.8 &       0.907 &     0.999 &   0.960 &    0.786 &       0.787 \\
	1.0   &    1.0 &       1.000 &     1.000 &   1.000 &    1.000 &       1.000 \\[0.2cm]
	\enddata
\end{deluxetable*}

It was necessary when constructing the mixing grid to fix $\log U_{\rm HII}$ (Section~\ref{sec:Method}), and we have tested the sensitivity of our results to this assumption by varying the fixed $\log U_{\rm HII}$ value.  The analysis presented in Figure~\ref{fig:fNLR_vs_mix_dist} and Figure~\ref{fig:BPT_NII_results} was performed for the three values $\log U_{\rm HII} = -3.50$, $-3.25$, and $-3.00$ with \Ep$\, = 45$\,eV.  In this experiment the lines of median \fN\ values of 0.2, 0.4 and 0.6 all changed by at most 0.07~dex in \OIIIHb.  Hence, we conclude that our results are insensitive to $\log U_{\rm HII}$, despite the low \fN\ values.

\vspace*{2.5cm}
\section{Conclusions} \label{sec:Conclusions}

Mixing between \HII\ region and AGN emission in SDSS galaxies is investigated by explicitly fitting `mixing' models to the data for the first time.  Our conclusions are as follows:
\begin{enumerate}
	\item We demonstrate that most Seyfert-classified spectra in SDSS have the majority of their Balmer line flux originating in \HII\ regions within the fixed fiber aperture, rather than in the NLR itself
	\item We show that our results are insensitive to the hardness of the assumed Seyfert ionizing continuum, and find that ionizing spectra peaking at energies of \Ep\,$ \sim 40 - 50$\,eV result in the most plausible distributions of inferred mixing fractions
	\item We provide equations of lines that quantify median `mixing levels' on the \NII\ BPT diagram
	\item We quantify the contributions to key emission-line fluxes from the model \HII\ and NLR spectra
\end{enumerate}

\acknowledgments
This research was conducted by the Australian Research Council Centre of Excellence for All Sky Astrophysics in 3 Dimensions (ASTRO 3D), through project number CE170100013.  B.G. acknowledges the support of the Australian Research Council as the recipient of a Future Fellowship (FT140101202).  M.D. and R.S. acknowledge support from ARC discovery project \#DP160103631.  This research is supported by an Australian Government Research Training Program (RTP) Scholarship.


\vspace*{1cm}
\begin{thebibliography}{}
	\expandafter\ifx\csname natexlab\endcsname\relax\def\natexlab#1{#1}\fi
	\providecommand{\url}[1]{\href{#1}{#1}}
	
	\bibitem[{{Abazajian} {et~al.}(2009){Abazajian}, {Adelman-McCarthy},
		{Ag{\"u}eros}, {Allam}, {Allende Prieto}, {An}, {Anderson}, {Anderson},
		{Annis}, {Bahcall}, \& et~al.}]{Abazajian_2009_SDSS7}3
	{Abazajian}, K.~N., {Adelman-McCarthy}, J.~K., {Ag{\"u}eros}, M.~A., {et~al.}
	2009, \apjs, 182, 543
	
	\bibitem[{{Baldwin} {et~al.}(1981){Baldwin}, {Phillips}, \&
		{Terlevich}}]{BPT1981}
	{Baldwin}, J.~A., {Phillips}, M.~M., \& {Terlevich}, R. 1981, \pasp, 93, 5
	
	\bibitem[{{B{\"a}r} {et~al.}(2017){B{\"a}r}, {Weigel}, {Sartori}, {Oh}, {Koss},
		\& {Schawinski}}]{Bar_2017_AGN_diagnostic}
	{B{\"a}r}, R.~E., {Weigel}, A.~K., {Sartori}, L.~F., {et~al.} 2017, \mnras,
	466, 2879
	
	\bibitem[{{Chen} {et~al.}(2013){Chen}, {Hickox}, {Alberts}, {Brodwin}, {Jones},
		{Murray}, {Alexander}, {Assef}, {Brown}, {Dey}, {Forman}, {Gorjian},
		{Goulding}, {Le Floc'h}, {Jannuzi}, {Mullaney}, \&
		{Pope}}]{Chen_2013_SFR_vs_BHAR}
	{Chen}, C.-T.~J., {Hickox}, R.~C., {Alberts}, S., {et~al.} 2013, \apj, 773, 3
	
	\bibitem[{{Constantin} {et~al.}(2008){Constantin}, {Hoyle}, \&
		{Vogeley}}]{Constantin_2008_AGN_in_voids}
	{Constantin}, A., {Hoyle}, F., \& {Vogeley}, M.~S. 2008, \apj, 673, 715
	
	\bibitem[{{Davies} {et~al.}(2014{\natexlab{a}}){Davies}, {Kewley}, {Ho}, \&
		{Dopita}}]{Davies_2014_Mixing_2}
	{Davies}, R.~L., {Kewley}, L.~J., {Ho}, I.-T., \& {Dopita}, M.~A.
	2014{\natexlab{a}}, \mnras, 444, 3961
	
	\bibitem[{{Davies} {et~al.}(2014{\natexlab{b}}){Davies}, {Rich}, {Kewley}, \&
		{Dopita}}]{Davies_2014_Mixing_1}
	{Davies}, R.~L., {Rich}, J.~A., {Kewley}, L.~J., \& {Dopita}, M.~A.
	2014{\natexlab{b}}, \mnras, 439, 3835
	
	\bibitem[{Dopita {et~al.}(2014)Dopita, Scharw\"{a}chter, Shastri, Kewley,
		Davies, Sutherland, Kharb, Jose, Hampton, Jin, Banfield, Basurah, \&
		Fischer}]{Dopita_2014_probing_ENLR_I_S7}
	Dopita, M.~A., Scharw\"{a}chter, J., Shastri, P., {et~al.} 2014, Astronomy \&
	Astrophysics, 566, A41.
	\newblock \url{http://adsabs.harvard.edu/abs/2014A\%26A...566A..41D}
	
	\bibitem[{{Greene} \& {Ho}(2005)}]{Greene_2005_SDSS_AGN_kinematics}
	{Greene}, J.~E., \& {Ho}, L.~C. 2005, \apj, 627, 721
	
	\bibitem[{{Greene} \& {Ho}(2007)}]{Greene_2007_SMBH_mass_function}
	---. 2007, \apj, 667, 131
	
	\bibitem[{{Jones} {et~al.}(2016){Jones}, {Hickox}, {Black}, {Hainline},
		{DiPompeo}, \& {Goulding}}]{Jones_2016_SDSS_LEdd}
	{Jones}, M.~L., {Hickox}, R.~C., {Black}, C.~S., {et~al.} 2016, \apj, 826, 12
	
	\bibitem[{{Kauffmann} \& {Heckman}(2009)}]{Kauffmann_2009_Edd_ratio}
	{Kauffmann}, G., \& {Heckman}, T.~M. 2009, \mnras, 397, 135
	
	\bibitem[{{Kauffmann} {et~al.}(2008){Kauffmann}, {Heckman}, \&
		{Best}}]{Kauffmann_2008_SDSS_radio_AGN}
	{Kauffmann}, G., {Heckman}, T.~M., \& {Best}, P.~N. 2008, \mnras, 384, 953
	
	\bibitem[{{Kauffmann} {et~al.}(2003){Kauffmann}, {Heckman}, {Tremonti},
		{Brinchmann}, {Charlot}, {White}, {Ridgway}, {Brinkmann}, {Fukugita}, {Hall},
		{Ivezi{\'c}}, {Richards}, \& {Schneider}}]{Kauffmann_2003}
	{Kauffmann}, G., {Heckman}, T.~M., {Tremonti}, C., {et~al.} 2003, \mnras, 346,
	1055
	
	\bibitem[{{Kewley} {et~al.}(2001{\natexlab{a}}){Kewley}, {Dopita},
		{Sutherland}, {Heisler}, \& {Trevena}}]{Kewley2001}
	{Kewley}, L.~J., {Dopita}, M.~A., {Sutherland}, R.~S., {Heisler}, C.~A., \&
	{Trevena}, J. 2001{\natexlab{a}}, \apj, 556, 121
	
	\bibitem[{{Kewley} {et~al.}(2006){Kewley}, {Groves}, {Kauffmann}, \&
		{Heckman}}]{Kewley2006_AGN_hosts}
	{Kewley}, L.~J., {Groves}, B., {Kauffmann}, G., \& {Heckman}, T. 2006, \mnras,
	372, 961
	
	\bibitem[{{Kewley} {et~al.}(2001{\natexlab{b}}){Kewley}, {Heisler}, {Dopita},
		\& {Lumsden}}]{Kewley_2001_optical_classification_IR_galaxies}
	{Kewley}, L.~J., {Heisler}, C.~A., {Dopita}, M.~A., \& {Lumsden}, S.
	2001{\natexlab{b}}, \apjs, 132, 37
	
	\bibitem[{{Kewley} {et~al.}(2005){Kewley}, {Jansen}, \&
		{Geller}}]{Kewley_2005_Aperture_effects}
	{Kewley}, L.~J., {Jansen}, R.~A., \& {Geller}, M.~J. 2005, \pasp, 117, 227
	
	\bibitem[{{Khabiboulline} {et~al.}(2014){Khabiboulline}, {Steinhardt},
		{Silverman}, {Ellison}, {Mendel}, \& {Patton}}]{Khabiboulline_2014_SDSS_AGN}
	{Khabiboulline}, E.~T., {Steinhardt}, C.~L., {Silverman}, J.~D., {et~al.} 2014,
	\apj, 795, 62
	
	\bibitem[{{Koss} {et~al.}(2011){Koss}, {Mushotzky}, {Veilleux}, {Winter},
		{Baumgartner}, {Tueller}, {Gehrels}, \& {Valencic}}]{2011_Koss_BAT_SDSS_AGN}
	{Koss}, M., {Mushotzky}, R., {Veilleux}, S., {et~al.} 2011, \apj, 739, 57
	
	\bibitem[{{Lagos} {et~al.}(2011){Lagos}, {Padilla}, {Strauss}, {Cora}, \&
		{Hao}}]{Lagos_2011_SDSS_AGN_alignment}
	{Lagos}, C.~D.~P., {Padilla}, N.~D., {Strauss}, M.~A., {Cora}, S.~A., \& {Hao},
	L. 2011, \mnras, 414, 2148
	
	\bibitem[{{Maragkoudakis} {et~al.}(2014){Maragkoudakis}, {Zezas}, {Ashby}, \&
		{Willner}}]{Maragkoudakis_2014_BPT_aperture_effects}
	{Maragkoudakis}, A., {Zezas}, A., {Ashby}, M.~L.~N., \& {Willner}, S.~P. 2014,
	\mnras, 441, 2296
	
	\bibitem[{{Mateos} {et~al.}(2013){Mateos}, {Alonso-Herrero}, {Carrera},
		{Blain}, {Severgnini}, {Caccianiga}, \& {Ruiz}}]{2013_Mateos_AGN}
	{Mateos}, S., {Alonso-Herrero}, A., {Carrera}, F.~J., {et~al.} 2013, \mnras,
	434, 941
	
	\bibitem[{{Miller} {et~al.}(2003){Miller}, {Nichol}, {G{\'o}mez}, {Hopkins}, \&
		{Bernardi}}]{Miller_2003_SDSS_AGN}
	{Miller}, C.~J., {Nichol}, R.~C., {G{\'o}mez}, P.~L., {Hopkins}, A.~M., \&
	{Bernardi}, M. 2003, \apj, 597, 142
	
	\bibitem[{{Paliya} {et~al.}(2018){Paliya}, {Ajello}, {Rakshit}, {Mandal},
		{Stalin}, {Kaur}, \& {Hartmann}}]{Paliya_2018_SDSS_AGN}
	{Paliya}, V.~S., {Ajello}, M., {Rakshit}, S., {et~al.} 2018, \apjl, 853, L2
	
	\bibitem[{{Reines} \& {Volonteri}(2015)}]{2015_Reines_BH_mass}
	{Reines}, A.~E., \& {Volonteri}, M. 2015, \apj, 813, 82
	
	\bibitem[{{Shirazi} \& {Brinchmann}(2012)}]{Shirazi_Brinchmann_2012_HeII}
	{Shirazi}, M., \& {Brinchmann}, J. 2012, \mnras, 421, 1043
	
	\bibitem[{{Sutherland} \&
		{Dopita}(2017)}]{Sutherland_Dopita_2017_shocks_MAPPINGSV}
	{Sutherland}, R.~S., \& {Dopita}, M.~A. 2017, \apjs, 229, 34
	
	\bibitem[{{Thomas} {et~al.}(2018){Thomas}, {Dopita}, {Kewley}, {Groves},
		{Sutherland}, {Hopkins}, \& {Blanc}}]{Thomas_2018_NebulaBayes}
	{Thomas}, A.~D., {Dopita}, M.~A., {Kewley}, L.~J., {et~al.} 2018, \apj, 856, 89
	
	\bibitem[{{Thomas} {et~al.}(2016){Thomas}, {Groves}, {Sutherland}, {Dopita},
		{Kewley}, \& {Jin}}]{Thomas_2016_oxaf}
	{Thomas}, A.~D., {Groves}, B.~A., {Sutherland}, R.~S., {et~al.} 2016, \apj,
	833, 266
	
	\bibitem[{{Tremonti} {et~al.}(2004){Tremonti}, {Heckman}, {Kauffmann},
		{Brinchmann}, {Charlot}, {White}, {Seibert}, {Peng}, {Schlegel}, {Uomoto},
		{Fukugita}, \& {Brinkmann}}]{Tremonti_2004_MZ}
	{Tremonti}, C.~A., {Heckman}, T.~M., {Kauffmann}, G., {et~al.} 2004, \apj, 613,
	898
	
	\bibitem[{{Veilleux} \& {Osterbrock}(1987)}]{Veilleux_1987_LineRatios}
	{Veilleux}, S., \& {Osterbrock}, D.~E. 1987, \apjs, 63, 295
	
	\bibitem[{{von der Linden} {et~al.}(2010){von der Linden}, {Wild}, {Kauffmann},
		{White}, \& {Weinmann}}]{von_der_Linden_2010_SDSS_clusters}
	{von der Linden}, A., {Wild}, V., {Kauffmann}, G., {White}, S.~D.~M., \&
	{Weinmann}, S. 2010, \mnras, 404, 1231
	
	\bibitem[{{Wild} {et~al.}(2010){Wild}, {Heckman}, \&
		{Charlot}}]{Wild_2010_SB_AGN_connection}
	{Wild}, V., {Heckman}, T., \& {Charlot}, S. 2010, \mnras, 405, 933
	
	\bibitem[{{York} {et~al.}(2000){York}, {Adelman}, {Anderson}, {Anderson},
		{Annis}, {Bahcall}, {Bakken}, {Barkhouser}, {Bastian}, {Berman}, {Boroski},
		{Bracker}, {Briegel}, {Briggs}, {Brinkmann}, {Brunner}, {Burles}, {Carey},
		{Carr}, {Castander}, {Chen}, {Colestock}, {Connolly}, {Crocker}, {Csabai},
		{Czarapata}, {Davis}, {Doi}, {Dombeck}, {Eisenstein}, {Ellman}, {Elms},
		{Evans}, {Fan}, {Federwitz}, {Fiscelli}, {Friedman}, {Frieman}, {Fukugita},
		{Gillespie}, {Gunn}, {Gurbani}, {de Haas}, {Haldeman}, {Harris}, {Hayes},
		{Heckman}, {Hennessy}, {Hindsley}, {Holm}, {Holmgren}, {Huang}, {Hull},
		{Husby}, {Ichikawa}, {Ichikawa}, {Ivezi{\'c}}, {Kent}, {Kim}, {Kinney},
		{Klaene}, {Kleinman}, {Kleinman}, {Knapp}, {Korienek}, {Kron}, {Kunszt},
		{Lamb}, {Lee}, {Leger}, {Limmongkol}, {Lindenmeyer}, {Long}, {Loomis},
		{Loveday}, {Lucinio}, {Lupton}, {MacKinnon}, {Mannery}, {Mantsch}, {Margon},
		{McGehee}, {McKay}, {Meiksin}, {Merelli}, {Monet}, {Munn}, {Narayanan},
		{Nash}, {Neilsen}, {Neswold}, {Newberg}, {Nichol}, {Nicinski}, {Nonino},
		{Okada}, {Okamura}, {Ostriker}, {Owen}, {Pauls}, {Peoples}, {Peterson},
		{Petravick}, {Pier}, {Pope}, {Pordes}, {Prosapio}, {Rechenmacher}, {Quinn},
		{Richards}, {Richmond}, {Rivetta}, {Rockosi}, {Ruthmansdorfer}, {Sandford},
		{Schlegel}, {Schneider}, {Sekiguchi}, {Sergey}, {Shimasaku}, {Siegmund},
		{Smee}, {Smith}, {Snedden}, {Stone}, {Stoughton}, {Strauss}, {Stubbs},
		{SubbaRao}, {Szalay}, {Szapudi}, {Szokoly}, {Thakar}, {Tremonti}, {Tucker},
		{Uomoto}, {Vanden Berk}, {Vogeley}, {Waddell}, {Wang}, {Watanabe},
		{Weinberg}, {Yanny}, {Yasuda}, \& {SDSS Collaboration}}]{York_2000_SDSS}
	{York}, D.~G., {Adelman}, J., {Anderson}, Jr., J.~E., {et~al.} 2000, \aj, 120,
	1579
	
\end{thebibliography}
\end{document}